\documentclass[a4paper,aps,preprintnumbers,showpacs,twocolumn,superscriptaddress,nofootinbib,amsmath,amssymb]{revtex4}
\usepackage{graphics}
\usepackage[utf8]{inputenc}
\usepackage[T1]{fontenc}
\usepackage{cmap}

\newcommand{\cf}{{cf.}~}

\newcommand{\eg}{{e.g.,}~}

\def\Order#1{{\cal O}\left(#1\right)}
\def\CarterC{{\cal C}}
\def\CarterK{{\cal K}}

\begin{document}
\title{Shadows of parametrized axially symmetric black holes \\ allowing for separation of variables}
\author{R. A. Konoplya}\email{roman.konoplya@gmail.com}
\affiliation{Research Centre for Theoretical Physics and Astrophysics, Institute of Physics, Silesian University in Opava, Bezručovo náměstí 13, CZ-74601 Opava, Czech Republic}
\affiliation{Peoples Friendship University of Russia (RUDN University), 6 Miklukho-Maklaya Street, Moscow 117198, Russian Federation}
\author{A. Zhidenko}\email{alexander.zhidenko@uni-oldenburg.de}
\affiliation{Institute of Physics, University of Oldenburg, D-26111 Oldenburg, Germany}
\affiliation{Centro de Matemática, Computação e Cognição (CMCC), Universidade Federal do ABC (UFABC),\\ Rua Abolição, CEP: 09210-180, Santo André, SP, Brazil}

\begin{abstract}
Metric of axially symmetric asymptotically flat black holes in an arbitrary metric theory of gravity can be represented in the general form which depends on infinite number of parameters. We constrain this general class of metrics by requiring the existence of additional symmetries, which lead to the separation of variables in the Hamilton-Jacobi and Klein-Gordon equations, and show that once the metric functions change sufficiently moderately in some region near the black hole, the black-hole shadow depends on a few deformation parameters only. We analyze the influence of these parameters on the black-hole shadow. We also show that the shadow of the rotating black hole in the Einstein-dilaton-Gauss-Bonnet theory is well approximated if the terms violating the separation of variables are neglected in the metric.
\end{abstract}
\maketitle

\section{Introduction}
Recent observations of shadows cast by the galactic black holes \cite{Goddi:2017pfy,Akiyama:2019cqa} open one more opportunity to test gravitational theory in the regime of strong field via probing the geometry of black holes. Adding extra fields of matter or modifying the gravitational theory may lead to deformation of the spacetime geometry in the vicinity of a black hole. Therefore, a great number of recent works were devoted to calculations of shadows of black holes in either alternative theories of gravity or in the presence of additional fields and matter (see \cite{Toshmatov:2021fgm,Khodadi:2020gns,Li:2021btf,Tsupko:2021yca,Cai:2021fpr,Chowdhuri:2020ipb,Qin:2020xzu,Jusufi:2020zln,Kala:2020prt,Rincon:2020cpz,He:2020dfo,Das:2020yxw,Konoplya:2020xam,Dokuchaev:2020wqk,Zeng:2020vsj,Khan:2020ngg,Badia:2020pnh,Ovgun:2020gjz,Neves:2020doc,Chen:2020qyp,Roy:2020dyy,Liu:2020ola,Konoplya:2020bxa,Vagnozzi:2020quf,Konoplya:2019xmn,Javed:2019rrg,Zhu:2019ura,Cunha:2016wzk,Mishra:2019trb,Lee:2021sws,Konoplya:2019sns} and references therein). Here we are interested in a more general approach to analysis of black-hole shadows, which is not linked to any particular theory of gravity, but instead uses the general parametrized description of rotating black holes in an arbitrary metric theory of gravity developed in \cite{Konoplya:2016jvv}. This parametrization is similar in spirit to the parametrized post-Newtonian formalism, which is valid not only far from the black hole, but in the whole space outside the event horizon. The parametrization includes, in its most general case, an infinite set of parameters, representing the expansion of a metric in the radial direction and around the equatorial plane. Afterward, this expansion, truncated at some order, was used in a number of papers for approximation of various numerical black-hole solutions \cite{Konoplya:2019goy,Konoplya:2019fpy,Kokkotas:2017ymc,Kokkotas:2017zwt,Hennigar:2018hza}. Further analysis of these approximate parametrized metrics for finding appropriate observable quantities \cite{Konoplya:2020jgt,Zinhailo:2019rwd,Konoplya:2019ppy,Konoplya:2019hml,Zinhailo:2018ska} showed that the convergence of the parametrization is superior due to the continued fraction expansion in the radial direction in terms of a compact coordinate. There is also the strict hierarchy of orders of expansion, so that one can always distinguish the dominant parameters. As the only general black-hole parametrization, it attracted considerable interest during the past few years and various effects were studied for such generalized description of black-hole spacetimes \cite{Franzin:2021kvj,Suvorov:2021amy,Kocherlakota:2020kyu,Volkel:2020daa,Konoplya:2021qll,Nampalliwar:2019iti,Yang:2020bpj,Volkel:2019muj}.

The natural question then is how to describe shadows for such a general axially symmetric and asymptotically flat black hole. For the class of metrics, which possesses sufficient symmetry allowing for the separation of variables in the Hamilton-Jacobi equation, there is a simple solution without resorting to the ray tracing and lengthy numerical computations. The subclass of such black-hole metrics (allowing also the separation of variables in the Klein-Gordon equation) and the corresponding general parametrization were studied in \cite{Konoplya:2018arm}. It was shown there that the above class of metrics can even be used as a good approximation for some black-hole solutions which do not allow for the separation of variables. Thus, if one discards the terms violating the separation of variables in the Einstein-dilaton-Gauss-Bonnet black hole \cite{Ayzenberg:2014aka}, observable quantities, such as quasinormal frequencies and binding energy of particles moving in the equatorial plane, change only insignificantly.
Following this line, for a number of characteristics (quasinormal modes, binding energy, frequency at the innermost stable circular orbit in the equatorial plane, and others) it was shown in \cite{Konoplya:2020hyk} that only a few parameters of deformation are important, provided the metric does not change too quickly in the region between the black-hole horizon and the so-called radiation zone. The latter was defined in \cite{Konoplya:2020hyk} as, conditionally, a region between the photon orbit and the innermost stable circular orbit. Dominant contributions in the astrophysically relevant effects of radiation come from this region. Such smooth behavior in the near-horizon region is what one could expect from a black-hole metric as an exact solution to some modified Einstein field equations, represented, for example, by the Kerr-Sen or Einstein-Weyl black holes, and what is opposite to rather artificial constructions of black-hole mimickers, possessing, for example, shells of matter near the surface of a compact body. In this context, particles' motion in such parametrized black-hole metrics was analyzed in \cite{Konoplya:2020hyk} and \cite{Konoplya:2018arm}, but only in the equatorial plane.

Having all the above in mind here we analyze shadows of asymptotically flat and axially symmetric black holes, given by the parametrized metric which possesses sufficient symmetry for the separation of variables. Apart from the fact that many black-hole metrics belong to this class, such a consideration may give the preliminary understanding of influence of the deformation parameters upon black hole's shadows in the most general case which must be treated with the ray-tracing method.

Here, by calculation of shadows, we further confirm the previous claim of \cite{Konoplya:2020hyk} that once the metric is \emph{moderate} in the sense suggested in \cite{Konoplya:2018arm}, it can be very well described by only a few deformation parameters. We will also show that the discarding of the terms violating the separation of variables in the metric functions of the Einstein-dilaton-Gauss-Bonnet black hole does not change the shadow seemingly.

The paper is organized as follows. In Sec.~\ref{sec:separability} we briefly review the class of black-hole metrics, allowing for the separation of variables in the Hamilton-Jacobi and Klein-Gordon equations. Sec.~\ref{sec:shadow} is devoted to the general formalism for analysis of circular orbits and shadows. In Sec.~\ref{sec:pars} we discuss the influence of various deformation parameters on the black-hole shadow.
In Sec.~\ref{sec:Kerr-Sen} we consider a particular example of Kerr-Sen black hole, while in Sec.~\ref{sec:EdGB} shadows cast by the Einstein-dilaton-Gauss-Bonnet black holes are analyzed.
In the Conclusions we summarize the obtained results and discuss open questions.

\section{Black-hole metric allowing for separation of variables}\label{sec:separability}
Supposing that the coordinates $r$ and $\theta$ are mutually orthogonal and orthogonal to the Killing coordinates $t$ and $\phi$, we can write the
line element of the general axially symmetric black hole \cite{Konoplya:2016jvv} as follows:
\begin{eqnarray}
ds^2 &=&
-\dfrac{N^2(r,\theta)-W^2(r,\theta)\sin^2\theta}{K^2(r,\theta)}dt^2
\nonumber \\&&
-2W(r,\theta)r\sin^2\theta dt \, d\phi
+K^2(r,\theta)r^2\sin^2\theta d\phi^2
\nonumber \\&&
+\Sigma(r,\theta)\left(\dfrac{B^2(r,\theta)}{N^2(r,\theta)}dr^2 +
r^2d\theta^2\right). \label{eq:initmetric}
\end{eqnarray}
Here the event horizon $r=r_0$ is determined by the solution to the equation $N^2(r_0, \theta) =0$.

The axially symmetric black-hole metric allowing for the separation of variables in the Hamilton-Jacobi and Klein-Gordon equations possesses a number of additional symmetries in comparison with the most general case. In particular, it has an additional Killing vector corresponding to the Carter constant \cite{Carter:1968ks}. In \cite{Konoplya:2018arm} it was shown that in this case the metric functions can be written in the following general form:
\begin{subequations}\label{eq:gen}
\begin{eqnarray}
B(r,\theta)&=&R_B(r),\label{eq:genB}\\
\Sigma(r,\theta)&=&R_\Sigma(r)+\dfrac{a^2\cos^2\theta}{r^2},\label{eq:gensigma}\\
W(r,\theta)&=&\dfrac{aR_M(r)}{r^2\Sigma(r,\theta)},\label{eq:genW}\\
N^2(r,\theta)&=&R_N(r)\equiv R_\Sigma(r)-\dfrac{R_M(r)}{r}+\frac{a^2}{r^2},\label{eq:genN}\\
K^2(r,\theta)&=&\dfrac{a W(r,\theta)}{r}+\dfrac{1}{\Sigma(r,\theta)}\Biggl(R_\Sigma^2(r)\label{eq:genK}\\\nonumber
&&+R_\Sigma(r)\dfrac{a^2}{r^2}+\dfrac{a^2\cos^2\theta}{r^2}N^2(r,\theta)\Biggr).
\end{eqnarray}
\end{subequations}

Due to the freedom to choose the radial coordinate, we can use the additional condition, $R_{\Sigma}(r)=1$, without loss of generality.

Finally, we introduce the compact radial coordinate
$$x\equiv1-\frac{r_0}{r},$$
and use the continued-fraction expansion in terms of $x$ in order to represent the other functions,
\begin{subequations}\label{parametrized}
\begin{eqnarray}
R_B&=&b_0(1-x)+\dfrac{b_1(1-x)^2}{1+\dfrac{b_2x}{1+\dfrac{b_3x}{1+\ldots}}}\,,\\
R_M&=&r_0\Biggr(1+\frac{a^2}{r_0^2}(1-x)^2+\epsilon x
\\\nonumber&&-(a_0-\epsilon)(1-x)x-\dfrac{a_1(1-x)^2x}{1+\dfrac{a_2x}{1+\dfrac{a_3x}{1+\ldots}}}\Biggr)\,,
\end{eqnarray}
\end{subequations}
where $\epsilon$ relates the horizon radius and the asymptotic mass $M$ in the following way
\begin{equation}\label{epsilon0}
\epsilon = \frac{2M-r_0}{r_0}\,.
\end{equation}
In the above expansion, there are two kinds of coefficients: those that are fixed by the asymptotic behavior of the metric and those that are determined by the near horizon properties. The coefficients $a_0$ and $b_0$ are of the first kind, that is, they are fixed by the asymptotic behavior and depend on the post-Newtonian parameters $\beta$ and $\gamma$ as follows:
\begin{eqnarray}
\label{a00}
a_0& = &(\beta-\gamma)\frac{2M^2}{r_0^2} =
\frac{(\beta-\gamma)(1+\epsilon)^2}{2}\,,
\\
\label{b00}
b_0& = &(\gamma-1)\frac{M}{r_0} = \frac{(\gamma-1)(1+\epsilon)}{2}\,.
\end{eqnarray}
On the contrary, the coefficients $a_1,a_2,a_3,\ldots$ and $b_1,b_2,b_3,\ldots$ describe the near-horizon geometry.

In \cite{Konoplya:2018arm} it was shown that the Hamilton-Jacobi equation,
\begin{equation}
p_{\mu}p^{\mu}=g^{\mu\nu}\dfrac{\partial S}{\partial x^\mu}\dfrac{\partial S}{\partial x^\nu}=-\mu^2,
\end{equation}
is separable for the metrics (\ref{eq:gen}).

After introducing the integrals of motion,
\begin{eqnarray}
\dfrac{\partial S}{\partial t}&=&E,\\
\dfrac{\partial S}{\partial \phi}&=&-L,
\end{eqnarray}
which are the total energy $E$ and the angular momentum $L$, we obtain
\begin{eqnarray}\label{HJeq}
&&\left(\dfrac{\partial S}{\partial \theta}\right)^2+\cot^2\theta L^2+\cos^2\theta a^2(\mu^2-E^2)
\\\nonumber&&
+\dfrac{r^2R_N(r)}{R_B^2(r)}\left(\dfrac{\partial S}{\partial r}\right)^2+\dfrac{R_\Sigma(r)}{R_N(r)}L^2+R_\Sigma(r)r^2(\mu^2-E^2)
\\\nonumber&&
-\dfrac{R_M(r)}{rR_N(r)}\left((L-aE)^2+R_\Sigma(r)r^2E^2\right)=0.
\end{eqnarray}

With the above equations at hand, we are ready to analyze geodesics in the black-hole spacetime.

\begin{figure*}
\resizebox{\linewidth}{!}{\includegraphics{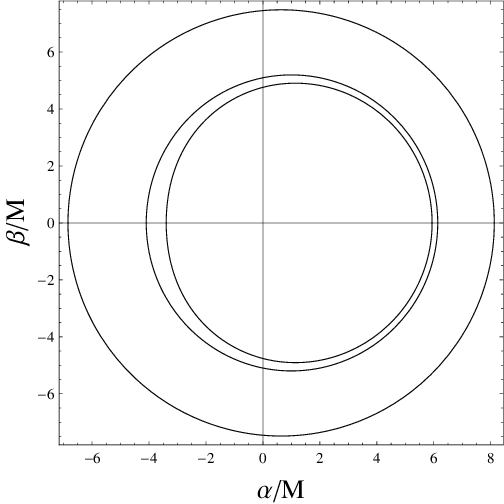}\includegraphics{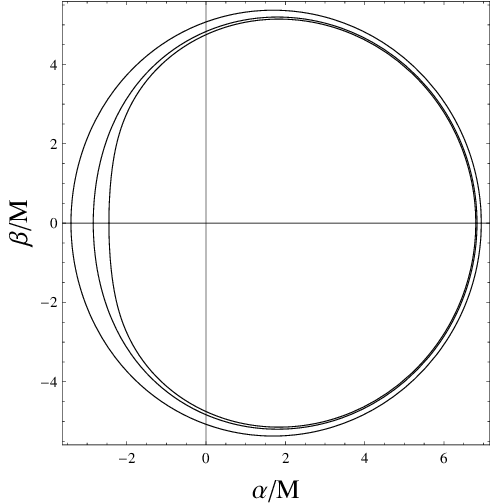}}
\caption{The profile of the shadow for $a=0.5$ (left) and $a=0.9$ (right). The inner profile is for $\epsilon = \epsilon_{Kerr} +0.25$ on the right plot ($\epsilon = \epsilon_{Kerr} +0.5$ on the left plot), the outer profile is for $\epsilon = \epsilon_{Kerr} -0.25$ on the right plot ($\epsilon = \epsilon_{Kerr} -0.5$ on the left plot). The middle profiles correspond to the Kerr shadows.}\label{fig:5}
\end{figure*}

\section{Circular orbits and the black-hole shadow}\label{sec:shadow}
We separate the angular and radial variables in (\ref{HJeq}) by introducing the Carter constant
\begin{equation}\label{Carter}
  \CarterC=\left(\dfrac{\partial S}{\partial \theta}\right)^2+\cot^2\theta L^2+\cos^2\theta a^2(\mu^2-E^2)>0,\\
\end{equation}
and find the master equation for the radial part
\begin{equation}\label{HJradial}
  \dfrac{r^4R_N^2(r)}{R_B^2(r)}\left(\dfrac{\partial S}{\partial r}\right)^2=V(r).
\end{equation}
Here the effective potential has the form
\begin{eqnarray}\label{V(r)}
V(r)&=&(r^2R_\Sigma(r)E+a^2E-aL)^2
\\\nonumber&&
-r^2R_N(r)(\CarterK+R_\Sigma(r)r^2\mu^2),
\end{eqnarray}
where
$$\CarterK\equiv\CarterC+(L-aE)^2>0.$$

The circular motion, corresponding to constant $r=r_c$, implies constant radial momentum and its first derivative. Hence it follows that
$$V(r_c)=0, \qquad V'(r_c)=0.$$
For the massless particles ($\mu=0$) the nontrivial solution to the above equations can be written in the following form (see, \eg\cite{Shaikh:2019fpu}):
\begin{eqnarray}
\frac{\CarterK}{E^2}&=&\frac{4\Delta(r_c)X'(r_c)^2}{\Delta'(r_c)^2},\\
\frac{aL}{E}&=&X(r_c)-\frac{2\Delta(r_c)X'(r_c)}{\Delta'(r_c)},
\end{eqnarray}
where we have introduced the two new functions
\begin{eqnarray}
X(r)&\equiv& r^2R_\Sigma(r)+a^2,\\
\Delta(r)&\equiv& r^2R_N(r)=X(r)-rR_M(r).
\end{eqnarray}

The circular orbit is unstable because $V''(r) >0$ there, and, once perturbed outward, a photon can leave the orbit and reach an observer located at a distance from the black hole. We consider an observer located at coordinates $(r,\theta,\phi)$. Due to the axial symmetry, we can fix $\phi=0$. The corresponding celestial coordinates are \cite{Vazquez:2003zm}
\begin{equation}\label{celestialdef}
\begin{array}{rcl}
\alpha&=&-r^2\sin\theta\dfrac{d\phi}{dr}=-r^2\sin\theta\dfrac{p^{\phi}}{p^{r}},\\
\beta&=&r^2\dfrac{d\theta}{dr}=r^2\dfrac{p^{\theta}}{p^r}.
\end{array}
\end{equation}

Substituting (\ref{Carter}) and (\ref{HJradial}) into (\ref{celestialdef}) and taking the limit $r\to\infty$, we can express the celestial coordinates for an asymptotic observer, $\alpha_a =\alpha(r.\theta)|_{r \to \infty}$ and $\beta_a= \beta(r,\theta)|_{r \to \infty}$, in terms of the constants of motion \cite{Vazquez:2003zm},
\begin{equation}\label{celestial}
\begin{array}{rrcll}
\alpha_a&=&\displaystyle\lim_{r\to\infty}\alpha &=&-\dfrac{L}{E \sin\theta},\\\mbox{}\\
\beta_a^2&=&\displaystyle\lim_{r\to\infty}\beta^2&=&\dfrac{\CarterC-\cot^2\theta L^2+\cos^2\theta a^2 E^2}{E^2},
\end{array}
\end{equation}

It is convenient to rewrite the above relations in the following form:
\begin{eqnarray}\nonumber
\alpha_a\cdot a\sin\theta&=&-\dfrac{aL}{E}=\dfrac{2\Delta(r_c)X'(r_c)}{\Delta'(r_c)}-X(r_c),
\\\label{shadoweqs}
\beta_a^2+(\alpha_a-a\sin\theta)^2&=&\frac{\CarterK}{E^2}=\frac{4\Delta(r_c)X'(r_c)^2}{\Delta'(r_c)^2}.
\end{eqnarray}

We notice that the celestial coordinates and, therefore, the shape of the black-hole shadow, do depend only on $R_M(r)$, that is, on the coefficients $\epsilon,a_0,a_1,a_2,a_3,\ldots$ and they do not depend on the rest of the coefficients $b_0,b_1,b_2,b_3,\ldots$.

The shape of the black-hole shadow is determined by the set of solutions to equations (\ref{shadoweqs}). In order to find the accurate shape of the black-hole shadow, we employ the polar celestial coordinates, $\rho$ and $\nu$,
$$\alpha_a=\rho\cos\nu+a\sin\theta, \qquad \beta_a=\rho\sin\nu,$$
then, for any $\nu\in[0,2\pi]$ we solve eqs.~(\ref{shadoweqs}) numerically and find the corresponding values of $\rho>0$ and $r_c>r_0$.

It is interesting to note, that, when $a\sin\theta=0$ the shadow is a circle, and the radius $\rho$ is given by the following relation:
\begin{equation}\label{minradius}
  \frac{1}{\rho^2}=\max\frac{\Delta(r)}{X^2(r)}=\max\frac{r^2R_N(r)}{(R_\Sigma(r)r^2+a^2)^2}.
\end{equation}
Therefore, the photon sphere is located in the vicinity of the maximum of the following function (\cf\cite{Konoplya:2020hyk}):
\begin{equation}\label{P(x)}
  P(x)\equiv\frac{(1-x)^2R_N(x)}{(R_\Sigma(x)+(1-x)^2a^2/r_0^2)^2}.
\end{equation}

The function $P(x)$ will further be important, because it allows us to see where the radiation zone starts. Considerable deviation of $P(x)$ from its Kerr values means correspondingly significant deviation of the black-hole shadow from its Kerr profile.

\section{Shadows of axially symmetric black holes: the role of the deformation parameters}\label{sec:pars}
The first question which we should address here is: supposing that the parametrization is used to approximate some black-hole metric, how many orders of the continued fraction expansion are sufficient to provide satisfactory accuracy of the approximation? As the main aim of any generic description of axially symmetric black holes is to study possible deviations from the Kerr geometry, we can define the \emph{effect} as the deviation of an observable quantity from its Kerr value. For example, such quantities could be the area of the shadow, frequency at the innermost stable circular orbit, or frequencies of quasinormal modes. Such observable quantity calculated for the accurate black-hole metric (given either numerically or in analytical form) will differ from those for the approximate metrics given by the parametrization in which the continued fraction expansion was truncated at some order. The difference between these values of the observable quantity for the accurate and approximate metrics is the \emph{error}. It is evident that once the effect is much larger than the error, the approximation is sufficient at least for the description of the effect.

The most influential parameter of deformation is $\epsilon$ which, roughly, determines the size of the black hole. Therefore, it is these parameters that must be most constrained by the current observations. We do not expect that it can change more than by a few tens of percents from its Kerr value,
$$\epsilon_{Kerr}\equiv\frac{a^2}{r_0^2}=\frac{a^2}{(M+\sqrt{M^2-a^2})^2},$$
and even this order of deviation should be considered as a stress test applicable only to measurements in the electromagnetic spectrum with the worst precision. In fig.~\ref{fig:5} we can see that the deviation of $\epsilon$ by $0.5$ leads to a noticeable change in the shadow size and should rather be considered as extremal.

\begin{figure}
\resizebox{\linewidth}{!}{\includegraphics{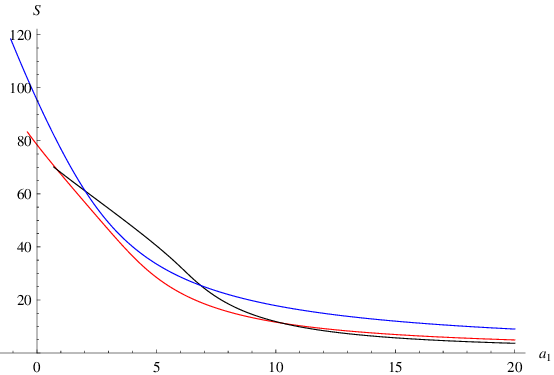}}
\caption{The area of the shadow as a function of $a_1$ when $a_2 = a_3 =...=0$. Here we have $\epsilon = \epsilon_{Kerr}$ (red), $\epsilon_{Kerr}+0.5$ (black, lower), and $\epsilon_{Kerr}-0.5$ (blue, upper). The values of the coefficient $a_1$ starts from its minimal allowed value.}\label{fig:1}
\end{figure}

\begin{figure}
\resizebox{\linewidth}{!}{\includegraphics{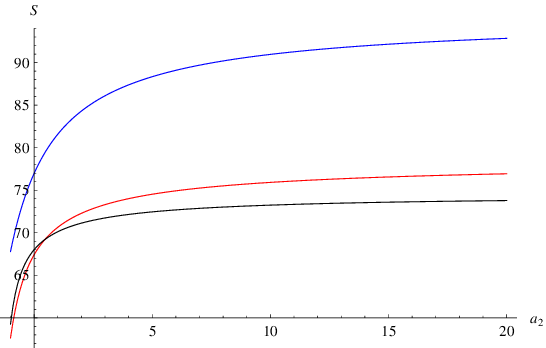}}
\caption{The area of the shadow as a function of $a_2$ when, $a_1=1$, $a_3 = a_4 =...=0$. Here we have $\epsilon = \epsilon_{Kerr}$ (red), $\epsilon_{Kerr}+0.5$ (black, lower), and $\epsilon_{Kerr}-0.5$ (blue, upper). The values of the coefficient $a_2$ starts from its minimal allowed value.}\label{fig:2}
\end{figure}

\begin{figure}
\resizebox{\linewidth}{!}{\includegraphics{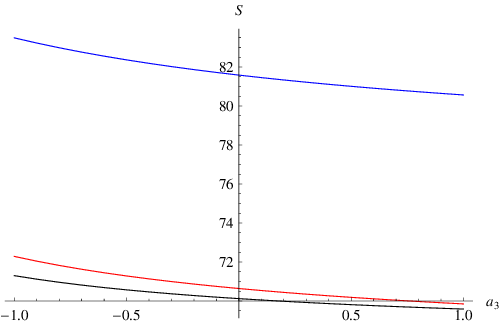}}
\caption{The area of the shadow as a function of $a_3$ when, $a_1=a_2 = 1$, $a_4 =...=0$. Here we have $\epsilon = \epsilon_{Kerr}$ (red), $\epsilon_{Kerr}+0.5$ (black, lower), and $\epsilon_{Kerr}-0.5$ (blue, upper).}\label{fig:3}
\end{figure}

\begin{figure}
\resizebox{\linewidth}{!}{\includegraphics{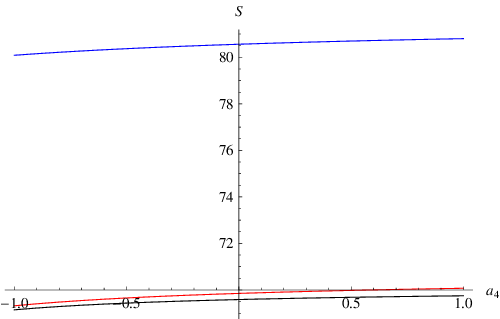}}
\caption{The area of the shadow as a function of $a_4$ when, $a_1=a_2 = a_3 = 1$, $a_5 =...=0$. Here we have $\epsilon = \epsilon_{Kerr}$ (red), $\epsilon_{Kerr}+0.5$ (black, lower), and $\epsilon_{Kerr}-0.5$ (blue, upper).}\label{fig:4}
\end{figure}

In fig.~\ref{fig:1} we can see the area of the shadow as a function of the parameters $a_1$ (for various values of $\epsilon$ and all the other null deformation parameters) which changes from its minimal value, compatible with the event horizon, until reaching the asymptotic regime, when the shadow practically does not change. The Kerr metric corresponds to $a_1=0$ of the red curve. Thus, we can see that the effect (deviation of the shadow area from its Kerr value) can reach quite few hundreds of percents. At the same time, once $a_1$ is fixed and the coefficient $a_2$ runs from its minimal (negative) value until the asymptotic regime, the maximal difference in the shadow's area is about tens of percents  (see fig.~\ref{fig:2}). The higher-order coefficients are even less significant, as can be seen in figs.~\ref{fig:3} and~\ref{fig:4}. Supposing that the second-order correction term, represented by the coefficient $a_2$, gives the order of the error for the first-order approximation, we can see that the rule requiring that the effect must be at least one order larger than the error is fulfilled. Thus, we conclude that already the first order of the expansion is sufficient to represent an axially symmetric black-hole metric. Then, the metric will be dependent only upon the three deformation parameters $\epsilon$, $a_1$, and $b_1$ (although the shadow does not depend on the latter coefficient),
\begin{eqnarray}\nonumber
ds^2&=&-\dfrac{R_N(r)-W^2(r,\theta)\sin^2\theta}{K^2(r,\theta)}dt^2
\\&& \nonumber
-2W(r,\theta)r\sin^2\theta dt \, d\phi
+K^2(r,\theta)r^2\sin^2\theta d\phi^2
\\&&
+\Sigma(r,\theta)\left(\dfrac{R_B^2(r)}{R_N(r)}dr^2 +
r^2d\theta^2\right),\label{metric}
\\\nonumber
R_N(r)&=&\left(1+\dfrac{a^2}{r^2}+\dfrac{r_0^3a_1}{r^3}\right)\left(1-\dfrac{r_0}{r}\right) - \dfrac{r_0\epsilon}{r} + \dfrac{r_0^3\epsilon}{r^3},\\ \nonumber
R_B(r)&=&1+\frac{r_0^2b_1}{r^2},\qquad \Sigma(r,\theta)=1+\frac{a^2}{r^2}\cos^2\theta,\\ \nonumber
W(r,\theta)&=&\frac{a}{r\Sigma(r,\theta)}\left(1+\frac{a^2}{r^2}-R_N(r)\right),\\ \nonumber
K^2(r,\theta)&=&\dfrac{r^2+a^2+a^2\cos^2\theta R_N(r)}{r^2\Sigma(r,\theta)}+\dfrac{a}{r}W(r,\theta).
\end{eqnarray}

\begin{figure}
\resizebox{\linewidth}{!}{\includegraphics{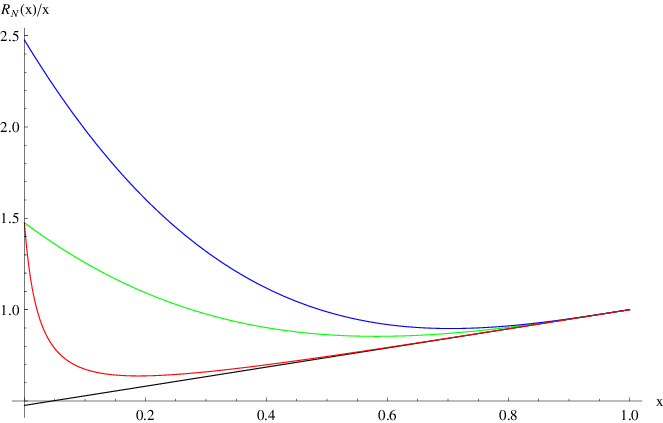}}
\resizebox{\linewidth}{!}{\includegraphics{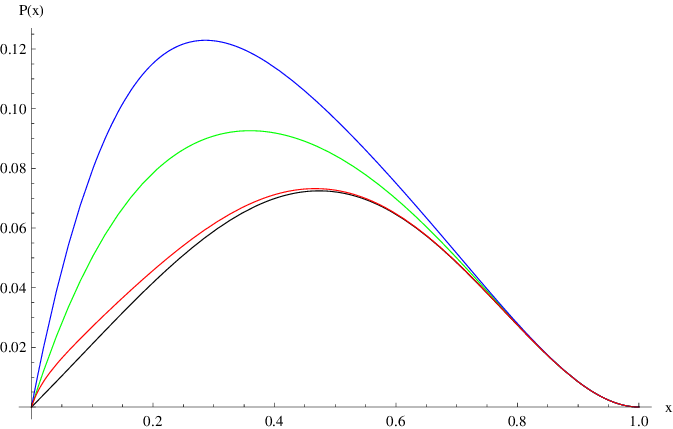}}
\caption{The $R_{N}(x)/x$ and $P(x)$ functions for the Kerr metric (black) and the three examples of deformed metrics (from lower to upper): (a) $a_1 =1$, $a_2 =40$ $a_3=a_4 =...=0$ (red); (b) $a_1 =1$, $a_2=a_3 =...=0$ (green); (c) $a_1=2$, $a_2=a_3 =...=0$ (blue). The parameters $\epsilon$ and $a_0$ are fixed by their Kerr values.}\label{fig:6}
\end{figure}

Another appealing question is how good the approximation given by the above general form of the metric would be for this or that black-hole solution?
One can imagine the situation in which the black-hole metric sharply changes in a small region in the vicinity of the event horizon, but remains Kerr-like in the rest of space. This happens, for instance, for black-hole mimickers. In some of such cases the parametrization may not be able to approximate the metric function properly in the whole space. At the same time, such mimicking of the Kerr solution would give no noticeable change in the observable quantities. Indeed, various radiation phenomena, be it quasinormal modes, gravitational lensing or other radiation in the electromagnetic spectrum are determined by the behavior of the metric functions in a region that starts roughly from the photon orbit and ends at the innermost stable circular orbit (ISCO). Therefore, we call this region \emph{radiation zone} \cite{Konoplya:2020hyk}. The behavior of the metric functions outside the radiation zone, that is, at a few distance from ISCO as well as in the near horizon zone is insignificant for scattering of waves and lensing. The metrics which change sufficiently smoothly between the event horizon and radiation zone we called \emph{moderate metrics} \cite{Konoplya:2020hyk}.

In Fig.~\ref{fig:6} we show examples of moderate (case b) and nonmoderate (cases a and c) metrics. The latter are characterized by significant change of the metric function between the horizon and the peak of the function $P(x)$ (radiation zone).
On examples of the equatorial-plane effects for particle motion and quasinormal modes considered in \cite{Konoplya:2020hyk} we can see that for moderate metrics the higher coefficients of the parametrization cannot be seemingly larger than the lower ones. As can be seen here on examples of black-holes shadows the same is true for the nonequatorial effects.
Thus, the above metric (\ref{metric}) can be used for description of moderate axially asymmetric and asymptotically flat black holes allowing for the separation of variables.

\section{Kerr-Sen black hole}\label{sec:Kerr-Sen}
\begin{table}
\begin{tabular}{|l|c|c|c|c|c|}
\hline
$a/\mu$&$0$&$1$&$2$&$4$&exact value \\
\hline
$0.5$   & 145.2708 & 145.2995 & 145.2954 & 145.2945 & 145.2945 \\
$0.95$  & 136.6061 & 136.6791 & 136.6687 & 136.6656 & 136.6656 \\
$0.998$ & 133.3219 & 133.4551 & 133.4379 & 133.4318 & 133.4319 \\
\hline
\end{tabular}
\caption{Shadow area ($\theta=\pi/2$) for different orders of approximation for the Kerr-Sen black holes ($b/\mu=0.5$) in units of $\mu^2$. The exact values are taken from \cite{Younsi:2016azx}.}\label{tabl:KerrSen}
\end{table}

A solution to the classical equations of motion arising in the low energy effective field theory for heterotic string theory has the solution corresponding to a charged rotating and asymptotically flat black hole. This black hole is described by the Kerr-Sen line element \cite{Sen:1992ua,Okai:1994td,Garcia:1995qz}. In our notations, this solution leads to the following expressions for the three metric functions:
\begin{equation}\label{Kerr-Sen}
\begin{array}{rcl}
  R_B(r)&=&1, \\
  R_M(r)&=& 2\mu+2b = 2M,\\
  R_\Sigma(r)&=&1+2b/r.
\end{array}
\end{equation}

The corresponding parametrization coefficients were found in \cite{Konoplya:2016jvv}:
\begin{subequations}\label{dilatonascoeff}
\begin{eqnarray}
\epsilon& = &\frac{2b+2\mu-r_0}{r_0}\,,\\
a_0& = &\frac{2b(b+\mu)}{r_0^2}\,,\\
a_1 &=& \frac{2(\mu+b)\left[2b^2+r_0^2+(2r_0-3b)\sqrt{r_0^2+b^2}\right]}
{r_0^2\sqrt{r_0^2+b^2}},\qquad \\
b_0 &=& 0\,,\\
b_1 &=& \frac{r_0}{\sqrt{r_0^2+b^2}}-1\,,\ldots
\end{eqnarray}
where the radius of the event horizon $r_0$ can be found from the following expression
\begin{equation}
r_0^2=\left(\mu + b + \sqrt{\mu^2 - a^2}\right)^2 - b^2.
\end{equation}
\end{subequations}

\begin{figure}
\resizebox{\linewidth}{!}{\includegraphics*{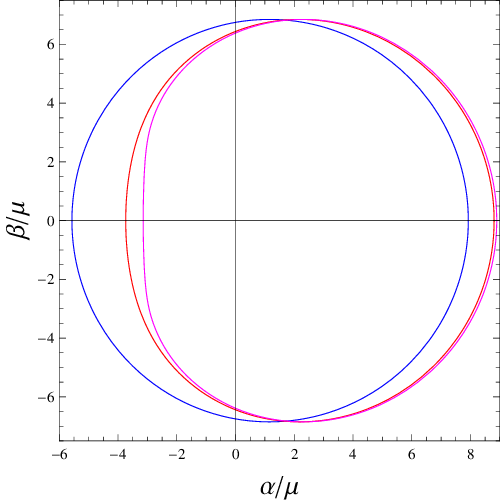}}
\caption{Shadows ($\theta=\pi/2$) the Kerr-Sen black holes ($b/\mu=0.5$): $a/\mu=0.5$ (blue), $a/\mu=0.95$ (red), and $a/\mu=0.998$ (magenta). }\label{fig:KerrSen}
\end{figure}

Applying the above parametrized approximation, truncated at various orders, to calculations of shadows, from Table~\ref{tabl:KerrSen} we can see that already the first-order approximation provides very good accuracy with a relative error below a small fraction of one percent. Higher orders increase the accuracy. The shape of the shadow for the Kerr-Sen black hole can be seen in fig.~\ref{fig:KerrSen}, where more prolate configurations correspond to higher rotation. It is worth mentioning that here we have also reproduced the shadows calculated in \cite{Younsi:2016azx} numerically by the ray-tracing method.

\section{Einstein-dilaton-Gauss-Bonnet black holes}\label{sec:EdGB}
Here we will consider a qualitatively different example of the black-hole metric, which does not allow for the separation of variables, represented by the solution to the Einstein-dilaton-Gauss-Bonnet equations \cite{Ayzenberg:2014aka}. Black holes in this theory have been studied in a great number of works, because they may provide quantum corrections to the black-hole geometry coming from the low energy limit string theory in the absence of the electromagnetic charge.
In \cite{Konoplya:2018arm} it was shown that, if we neglect the terms violating the separation of variables, then the slowly rotating Einstein-dilaton-Gauss-Bonnet black hole has the following metric function $R_M(r)$:
\begin{eqnarray}\nonumber
&&R_M(r) = 2 M -\zeta\Biggl(\frac{M^3}{3 r^2} + \frac{26 M^4}{3 r^3} + \frac{22 M^5}{5
  r^4} + \frac{32 M^6}{5 r^5}
\\&& \quad \nonumber
- \frac{80 M^7}{3 r^6}\Biggr)
+\zeta\frac{a^2}{r^2}\Biggl(\frac{3267 M}{1750 r} +
\frac{5017M^2}{875 r^2} + \frac{136819 M^3}{18375 r^3}
\\&& \quad
+ \frac{35198 M^4}{18375 r^4} - \frac{3818 M^5}{735 r^5} - \frac{4504 M^6}{245 r^6} + \frac{16 M^7}{5 r^7}\Biggr)\,,
\label{RMEdGB}
\end{eqnarray}
where we have included the terms depending on the coupling constant $\zeta$ of the order $\Order{\zeta}$ and $\Order{\zeta a^2}$.

Since the function $P(x)$ (\ref{P(x)}) depends on $a^2$, the dominant contribution to the shadow size for $\theta=0$ is
\begin{equation}\label{rhoGB}
  \rho^2=\frac{(r_s^2+a^2)^2}{r_s^2R_N(r_s)}+\Order{a^2},
\end{equation}
where $r_s$ corresponds to the radial coordinate of the null circular orbit for the nonrotating black hole, satisfying,
\begin{equation}\label{rsGB}
\dfrac{2\Delta(r_s)X'(r_s)}{\Delta'(r_s)}-X(r_s)\Biggr|_{a=0}=0.
\end{equation}

From (\ref{rsGB}) we find that
\begin{equation}\label{rsexp}
r_s=3M-\zeta\frac{961M}{810}+\Order{\zeta^2}.
\end{equation}
Substituting (\ref{rsGB}) into (\ref{rhoGB}), we obtain for $\theta=0$,
$$\rho^2=27M^2-3a^2-\zeta\frac{4397 M^2}{405}+\zeta\frac{3199459a^2}{357210}+\Order{\zeta^2,a^3}.$$
Therefore, the shadow area is given by the following relation:
\begin{equation}\label{GBarea}
S\approx S_{Kerr}\left(1-\zeta\frac{4397}{10935}+\zeta\frac{102539a^2}{357210M^2}\right),
\end{equation}
where $S_{Kerr}$ is the area of the shadow of the Kerr black hole with the same rotation parameter. The formula (\ref{GBarea}) is accurate for $\theta=0$ up to the order $\Order{\zeta^2,a^3}$, yet we can use it for any $\theta$ as an approximation, neglecting a deformation of the order $\Order{\zeta\cdot a\cdot\sin\theta}$.

\begin{table}
\begin{tabular}{|l|c|c|}
\hline
Shadow area ($M^2$ units)&$\zeta=0.1$&$\zeta=0.15$\\
\hline
Accurate & 79.977 & 77.853 \\
\hline
Second-order approximation & 80.689 & 78.983 \\
Error & 0.89\% & 1.45\%\\
\hline
Separable variables & 80.575 & 78.790\\
Error & 0.75\% & 1.20\%\\
\hline
Analytic formula & 80.801 & 79.421 \\
Error & 1.03\% & 2.01\%\\
\hline
\end{tabular}
\caption{Comparison of the accurate shadow area for the rotating Einstein-dilaton-Gauss-Bonnet black holes ($a=0.5M$) and the area computed using the second-order expansion in polar coordinate in \cite{Younsi:2016azx} with the approximation by the metric without terms violating the separation of variables and the approximation given by the analytic formula (\ref{GBarea}). The effect is calculated by comparing with the Kerr value of the area $S_{Kerr}\approx83.561M^2$. }\label{tabl:EdGB}
\end{table}

An important observation which can be done from the data presented in Table~\ref{tabl:EdGB} is that the area of the shadow changes insignificantly when the term violating the separation of variables is discarded. The relative error is of the same order as for the approximation by the second-order expansion in $\cos\theta$ \cite{Younsi:2016azx}, providing  sufficient accuracy to tell the shadow of the Einstein-dilaton-Gauss-Bonnet black hole from the Kerr one. Thus, we show that the described here class of black holes (given by eqs.~\ref{eq:initmetric}, \ref{eq:gen}, \ref{parametrized}) could be an effective model even for the solutions which do not allow for separation of variables. Thereby, we confirm the observations made in \cite{Konoplya:2018arm} where motion of particles was considered only in the equatorial plane.

\section{Conclusions}
The general parametrized form of the axially symmetric black-hole metrics is useful for description of various black-hole properties \cite{Konoplya:2021qll,Nampalliwar:2019iti}.
Here we considered shadows cast by the asymptotically flat and axially symmetric black holes whose spacetimes admit the Carter constant and therefore the geodesic equations allow for the separation of variables. Calculation of shadows in this case can be done without usage of the numerical ray-tracing method. We have shown the following:
\begin{itemize}
\item There is the strict hierarchy of the parameters of deformation as to the influence on the black-hole shadows. The dominant parameter which robustly determines the black-hole shadow is $\epsilon$. The next correction parameter $a_1$ is sufficient to achieve the regime in which the effect is much larger than the error due to truncation of the rest of deformation parameters.
\item  The situations in which we are guaranteed that the continued fraction series will converge quickly, that is, only the first few orders of expansion are enough to represent the black-hole geometry, are expressed by the requirement that the metric functions do not change quickly in a region from the event horizon until the end of the radiation zone. Such metrics are called \emph{moderate} and we present the compact general form of such a metric (eq.~\ref{metric}).
\item The presented general parametrized form of the metric allowing for the separation of variables in the Hamilton-Jacobi equation can also be effective for description of shadows of black-hole geometries which do not allow for such separation, as was shown here on the example of the black holes in the Einstein-dilaton-Gauss-Bonnet theory.
\end{itemize}

It is tempting to extend our work to the most general case of the axially symmetric black holes without any additional symmetries and see the role of all the possible parameters of the deformation with the help of the numerical ray tracing \cite{progress}. This would provide the constrains on the values of the parameters of deformation and show how close are the current observations in the electromagnetic spectrum to the Kerr geometry. The present work would give then a firm basis for checking such purely numerical results and provide already general understanding on the role of some of the parameters of deformation.

The class of black holes allowing for the separation of variables could also be further extended in the spirit of \cite{Papadopoulos:2020kxu} and the asymmetry relatively the equatorial plane could be included \cite{Chen:2020aix} as well.

\begin{acknowledgments}
R.~A.~K. acknowledges the support of the grant 19-03950S of Czech Science Foundation (GAČR).
The work of A.~Z. was supported by the Alexander von Humboldt Foundation, Germany.
\end{acknowledgments}

\end{document}